\documentclass[pre,twocolumn,showpacs,floatfix,superscriptaddress]{revtex4}
\usepackage{graphicx,amsmath,amssymb,url}

\begin{document}
\title{Majority-vote model on hyperbolic lattices}
\author{Zhi-Xi Wu}
\email{zhi-xi.wu@physics.umu.se} \affiliation{Department of
Physics, Ume{\aa}\, University, 901 87 Ume{\aa}, Sweden}
\author{Petter Holme}
\email{petter.holme@physics.umu.se}\affiliation{Department of
Physics, Ume{\aa}\, University, 901 87 Ume{\aa}, Sweden}
\affiliation{Department of Energy Science, Sungkyunkwan
University, Suwon 440--746, Korea}

\begin{abstract}
We study the critical properties of a non-equilibrium statistical
model, the majority-vote model, on heptagonal and dual heptagonal
lattices. Such lattices have the special feature that they only
can be embedded in negatively curved surfaces. We find, by using
Monte Carlo simulations and finite-size analysis, that the
critical exponents $1/\nu$, $\beta/\nu$ and $\gamma/\nu$ are
different from those of the majority-vote model on regular
lattices with periodic boundary condition, which belongs to the
same universality class as the equilibrium Ising model. The
exponents are also from those of the Ising model on a hyperbolic
lattice. We argue that the disagreement is caused by the effective
dimensionality of the hyperbolic lattices. By comparative studies,
we find that the critical exponents of the majority-vote model on
hyperbolic lattices satisfy the hyperscaling relation
$2\beta/\nu+\gamma/\nu=D_{\mathrm{eff}}$, where $D_{\mathrm{eff}}$
is an effective dimension of the lattice. We also investigate the
effect of boundary nodes on the ordering process of the model.
\end{abstract}
 \maketitle

\section{Introduction}\label{intro}
Recently, there has been a growing interest  the critical behavior
of statistical-physics models on curved surfaces---ranging from
spin models, such as the ferromagnetic Ising
model~\cite{Shima2006jpa,Shima2006jsm,Krcmar2008pre,Sakaniwa2009pre},
the \emph{XY} model~\cite{Baek2007epl,Baek2009prea}, the
Heisenberg model~\cite{Moura2007pla}, the \emph{q}-state clock
models~\cite{Gendiar2008pre,Baek2009preb}, to other traditional
models, such as percolation~\cite{Baek2009prec},
diffusion~\cite{Baek2008pre}, etc. One reason for this interest is
that many newly discovered soft materials (e.g., carbon nanotubes)
show a negatively curved structure in the
nanoscale~\cite{Park2003prl}. One peculiar feature of negatively
curved surfaces is that their boundary is a finite fraction of the
whole geometry~\cite{Coxeter1969book}. This structure has been
verified having a nontrivial impact on the critical behavior of
many models of statistical physics. For example, in the context of
the Ising model, significant shifts in static and dynamic critical
exponents toward the mean-field values were
noticed~\cite{Shima2006jpa,Shima2006jsm}; small-sized
ferromagnetic domains were observed to exist at temperatures far
greater than the critical temperature~\cite{Sakaniwa2009pre}; An
apparent zero-temperature orientational glass transition in the
\emph{XY} spin model on a negatively curved surface was recently
demonstrated~\cite{Baek2009prea}.

These findings motivate us to investigate another spin models on
negatively curved surfaces. The majority-vote
model~\cite{Oliveira1992jsp,Oliveira1993jpa,Marques1993jpa,Santos1995jsp}
is  a simple non-equilibrium model exhibiting up-down symmetry
that has been argued to belong to the universality class of the
equilibrium Ising model~\cite{Grinstein1985prl}. Oliveira first
verified this conjecture  on a square lattice with periodic
boundary conditions (i.e., a torus)~\cite{Oliveira1992jsp}.
Subsequently, the majority-vote model has been investigated on
regular lattices (with dimension larger than
two)~\cite{Marques1993jpa, Santos1995jsp,Kwak2007pre,Yang2008pre},
random lattice~\cite{Lima2005pre}, directed or undirected random
graphs~\cite{Pereira2005pre,Lima2009pa}, small world
networks~\cite{Campos2003pre}, and scale-free
networks~\cite{Lima2006ijmpc}, etc. Very recently, it has been
found that the critical behavior of the majority-vote model on
square lattice is also independent of transition rates (e.g., the
Glauber or Metropolis rates)~\cite{Kwak2007pre}. It has also been
observed that the majority-vote models defined on different
complex networks belong to different universality
classes~\cite{Pereira2005pre,Lima2009pa,Campos2003pre,Lima2006ijmpc}.

Our goal in this contribution is therefore to identify the
critical behavior of the majority-vote model when the underlying
lattice is embedded in a hyperbolic surface, in particular the
heptagonal and the dual heptagonal lattices, and investigate if
such an interaction structure is capable of modifying the critical
exponents. To this end, we use Monte Carlo (MC) simulations and
standard finite-size scaling techniques to determine the critical
noise parameter $q_c$ (the main control parameter of the
majority-vote model, as well as the critical exponents $1/\nu$,
$\beta/\nu$ and $\gamma/\nu$. Our numerical results suggest that
the critical exponents, in the stationary state, are different
from those of the Ising model confined to regular and hyperbolic
lattices.

In the following Sec.~\ref{model}, we  define our model, describe
the quantities we measure and the computational details. In
Sec.~\ref{results}, we present our numerical results and analysis.
Finally, we summarize and contextualize the observations in
Sec.~\ref{conclusion}.

\section{Model and simulation}\label{model}
\subsection{Majority-vote model}\label{mv}
Following
Refs.~\cite{Oliveira1992jsp,Oliveira1993jpa,Marques1993jpa,Santos1995jsp,
Lima2005pre,Campos2003pre,Lima2009pa,Pereira2005pre}, the
majority-vote model with noise is defined by a set of spin
variables \{$\sigma_{i}$\}, where each spin is associated to one
node of the heptagonal lattice and can take the values $\pm 1$.
The system evolves as follows: For each spin $i$, we first
determine the majority spin of $i$'s neighborhood. With
probability $q$ the $i$ takes the opposite sign of the majority
spin, otherwise it takes the same spin as the majority spin. The
probability $q$ is called the noise parameter and plays the same
role of temperature in equilibrium spin systems. In terms of $q$,
the probability of a single-spin-flip is given by
\begin{equation}\label{prob}
w (\sigma _{i})=\frac{1}{2}\left[1-(1-2q)\sigma_{i}
S\left(\sum_{j\in\Omega_i}\sigma _{j}\right)\right],
\end{equation}
where $S(x)=\mbox{sgn}(x)$ if $x \neq 0$ and $S(0)=0$, and the
summation is over all the neighboring spins of the focal site $i$.
The transition probability~(\ref{prob}) satisfies up-down symmetry
under the change of signs of the spins in the neighborhood of $i$.
In the limit of zero noise, the majority-vote model is
identical to the ferromagnetic Ising model at zero
temperature~\cite{Oliveira1992jsp,Campos2003pre}.

\subsection{Heptagonal lattice and dual heptagonal lattice}\label{hyperbolic}
\begin{figure}[h]
\includegraphics[width=0.32\linewidth]{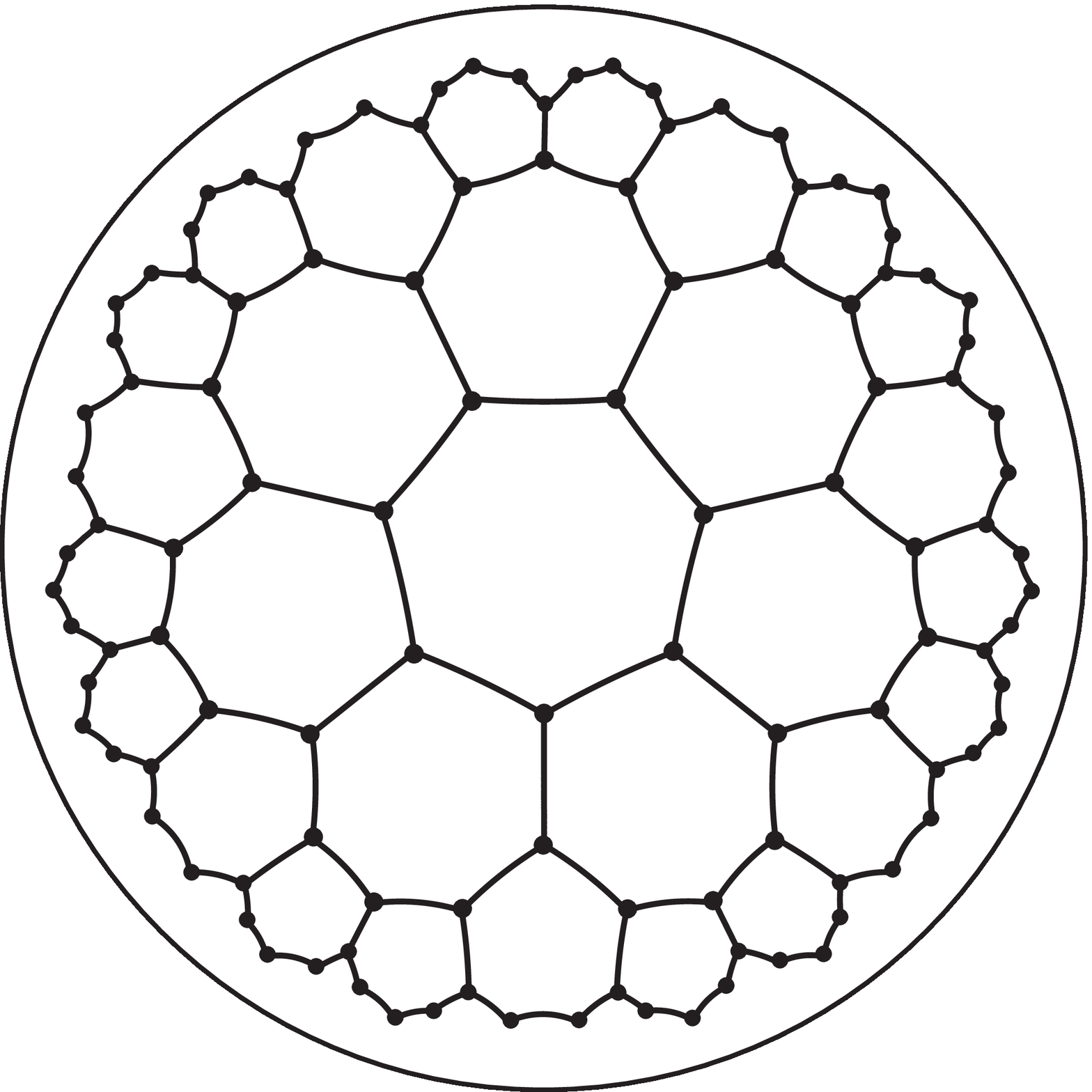}
\includegraphics[width=0.5\linewidth]{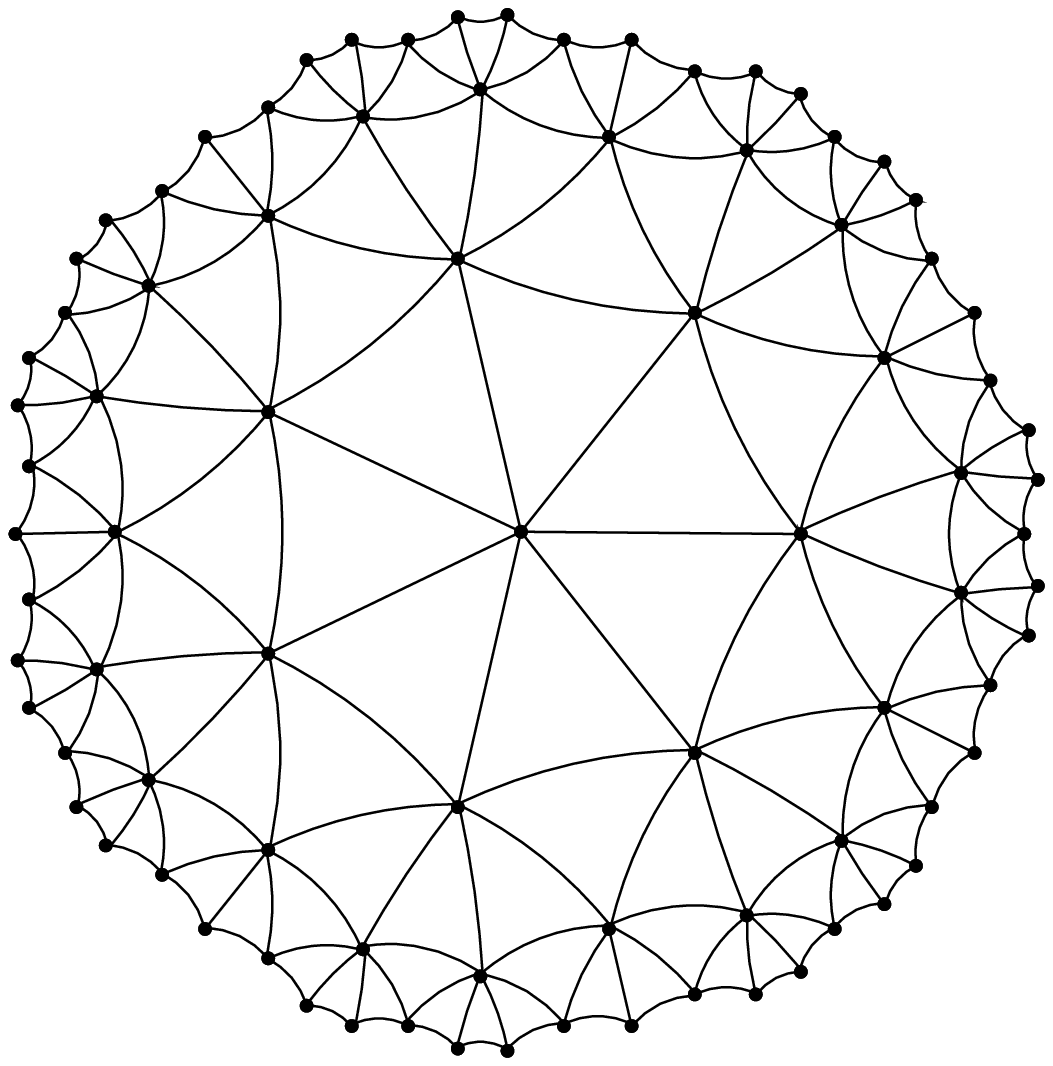}
\caption{A heptagonal lattice with level $3$ (left), and the dual
lattice with level $4$ (right). For the heptagonal lattices with
level $l=2, 3, 4, 5, 6, 7$, the total number of nodes are $35,
112, 315, 847, 2240, 5887$, respectively. For the dual lattice with
level $l=3, 4, 5, 6, 7, 8$, the total number of nodes are $29, 85,
232, 617, 1625, 4264$, respectively.}\label{HL}
\end{figure}

Figure~\ref{HL} shows two examples of the heptagonal lattice and
the dual heptagonal lattice. One peculiar property of the
heptagonal lattice is that, if we consider the innermost heptagon
as the level one, then the number of nodes of a heptagonal lattice
with level $l$ can be calculated by using the formulation
\cite{Sakaniwa2009pre,Baek2009preb}, $N(l) = 1 +
\frac{15}{\sqrt{5}} \sum_{j=1}^{l} [(\frac{3 + \sqrt{5}}{2})^j -
(\frac{3 - \sqrt{5}}{2})^j]$, which grows exponentially with level
$l$. In other words, the ratio of the perimeter to the area of the
lattice remains finite (about $0.62$) in the thermodynamic limit
$l\to\infty$~\cite{Sakaniwa2009pre}. If we make map each heptagon
in the heptagonal lattice to a node, and put a link between
adjacent heptagons, then we get the \textit{dual heptagonal
lattice} [Fig.~\ref{HL}(b)]. Since these lattices can only be
embedded in a hyperbolic surface with a constant negative
curvature, they are also called hyperbolic lattices.

\subsection{Measurements and finite size scaling}\label{measure}
In order to study the critical behavior of the model we consider
the magnetization $M$, the susceptibility $\chi$, and the Binder's
fourth-order cumulant $U$. These quantities are defined as
follows:
\begin{equation}
M=\left\langle\left\langle m\right\rangle _{T}\right\rangle_{C}=
\left\langle\left\langle\frac{1}{N}\left|\sum_{1}^{N}\sigma_{i}
\right|\right\rangle_{T}\right\rangle_{C},
\end{equation}
\begin{equation}
\chi=N\left[\langle~\langle m^{2}\rangle_{T}\rangle_{C}-\langle~
\langle m\rangle_{T}\rangle_{C}^{2}\right],
\end{equation}
\begin{equation}
U=1-\frac{\langle~\langle m^4\rangle_{T}\rangle_{C}}{3\langle~
\langle m^2\rangle_{T}\rangle_{C}^2},
\end{equation}
where $N$ is the total number of nodes of the hyperbolic lattice,
$\langle ...\rangle_{T}$ denotes thermodynamics averages taken in
the stationary regime, and $\langle \dots\rangle_{C}$ stands for
configurational averages.

The above quantities are functions of $q$ and $N$, in the critical
region, we expect the following finite-size scaling relations
\begin{equation}
M(N,q)=N^{-\beta/\nu}\tilde{M}(N^{1/\nu}\varepsilon)
\end{equation}
\begin{equation}
\chi(N,q)=N^{\gamma/\nu}\tilde{\chi}(N^{1/\nu}\varepsilon)
\end{equation}
\begin{equation}
U^\prime(N,q)=N^{1/\nu}\tilde{U}^\prime(N^{1/\nu}\varepsilon)
\end{equation}
where $\varepsilon=q-q_{c}$ and $U^\prime$ is the derivative of
Binder's fourth-order cumulant with respect to the noise. By the
standard finite-size scaling approach~\cite{Privman1990book}, we
assume scaling functions $\tilde{M}$, $\tilde{\chi}$, and
$\tilde{U}$ that are continuous and differentiable  in the
vicinity of the critical noise  $q_c$. From the size dependence of
$M$ and $\chi$ we can obtain the exponents $\beta/\nu$ and
$\gamma/\nu$, respectively. One alternative way to detect
$\gamma/\nu$,  since it also scales as
$N^{\gamma/\nu}$~\cite{Privman1990book}, is by plotting the
maximum value of the susceptibility versus $N$.

\subsection{Computational method}
We implement our MC simulations on the hyperbolic lattices with
various system sizes starting with all spins up and going from low
noise to high noise. It has been pointed out that this method can
reduce  the relaxation time considerably, especially in the low
noise limit, compared to starting with spins randomly oriented up
or down ~\cite{Stauffer2008jsm}. For each given $q$, we simulated
systems of size $N=112, 315, 847, 2240, 5887$ for the heptagonal
lattice, and $N=85, 232, 617, 1625, 4264$ for the dual heptagonal
lattice. For the sake of comparison, we also studied the
majority-vote on square lattice with free boundary condition. In
all simulations, we first wait $90\,000$ MC steps to let the
system attain stationary state (in the high noise case, this
number reduces to $50\,000$), and then ran another $30\,000$ MC
steps to get the average values. One MC step contains a sweep of
the spins in a random sequence. After every tenth MC steps, we
reshuffle the random sequence. Each data point presented below are
averages over $500$, $300$ and $200$ trials for $N<1000$,
$1000<N<4000$, and $N>4000$, respectively.

\begin{figure*}
\includegraphics[width=0.3\linewidth]{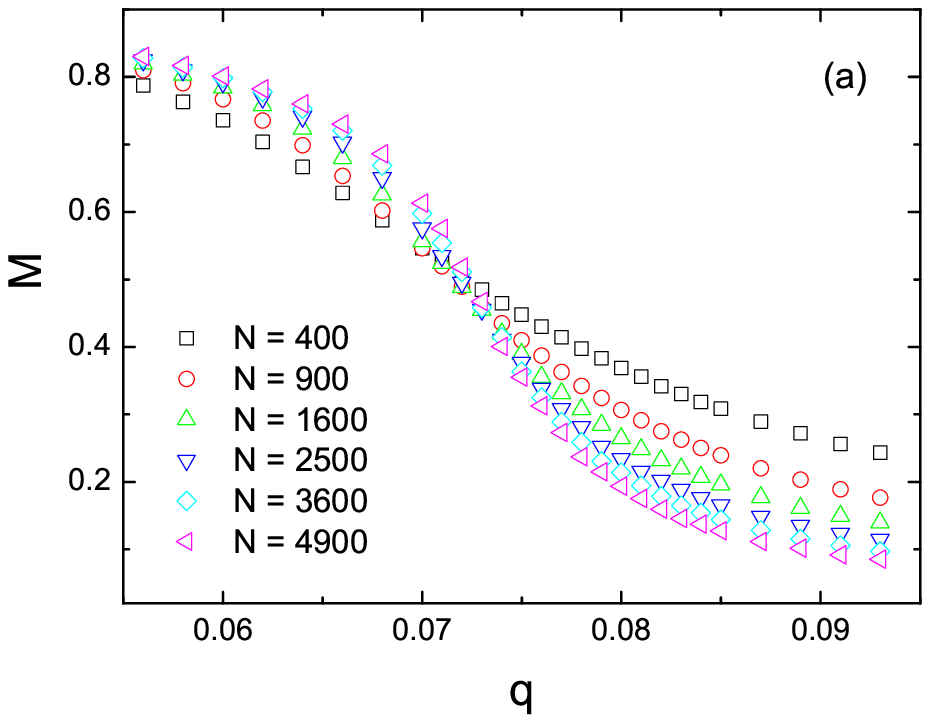}
\includegraphics[width=0.3\linewidth]{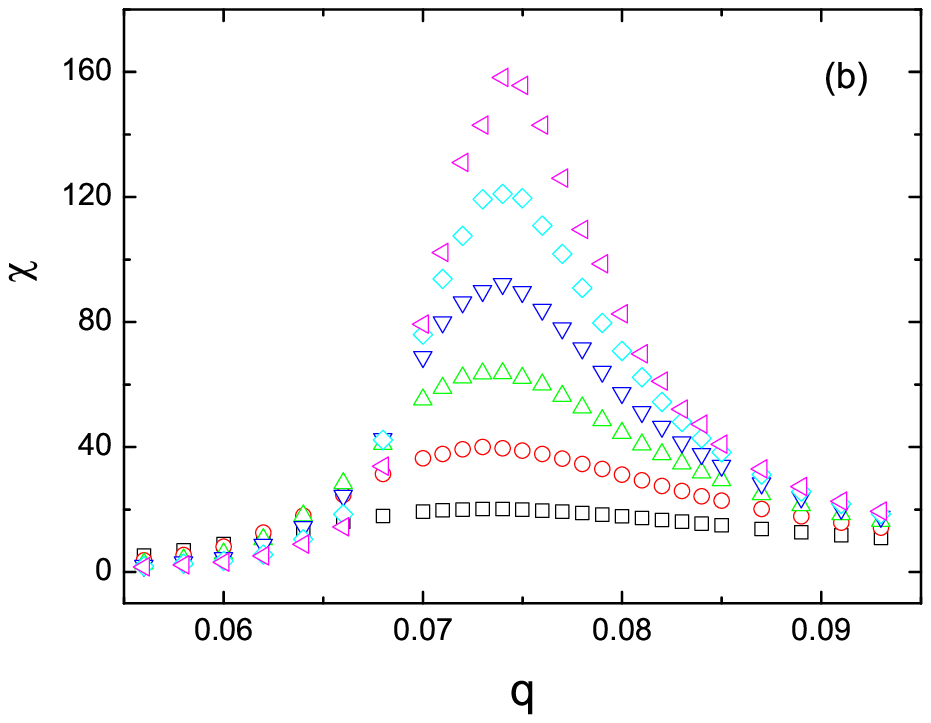}
\includegraphics[width=0.3\linewidth]{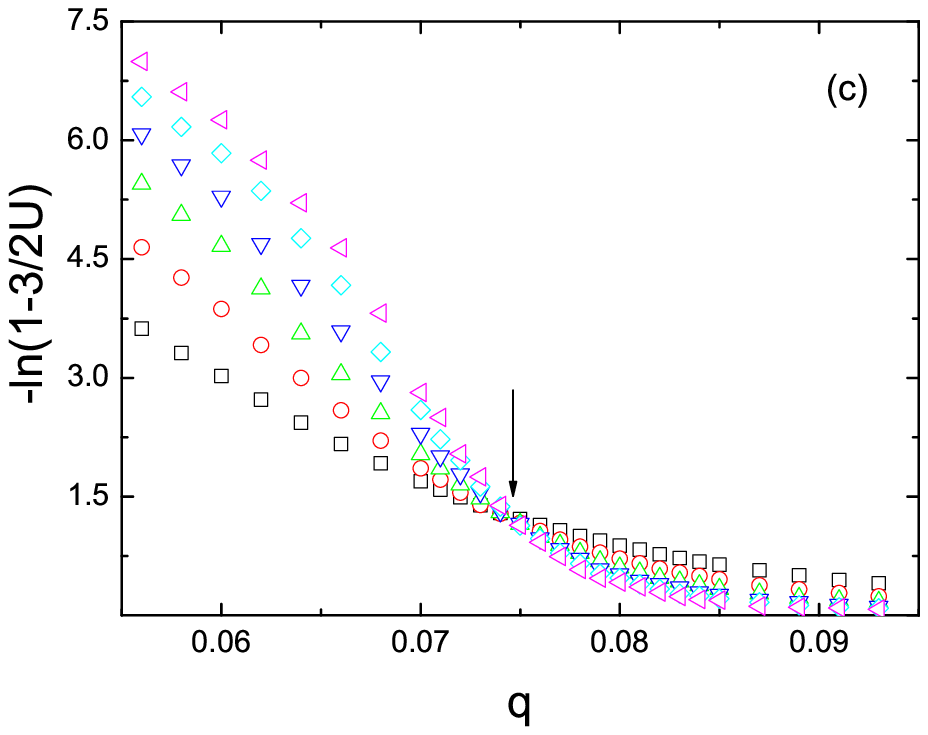}
\caption{(Color online) Majority-vote model on a two dimensional
square lattice with a free boundary condition. Magnetization $M$
(a), susceptibility $\chi$ (b), and reduced fourth-order cumulant
$U$ (c), as a function of the noise parameter $q$ for several
values of the system size $N$. In (c), within the accuracy of the
data, all curves intersect at $q_c=0.074(5)$.}\label{MVS}
\end{figure*}

\begin{figure*}
\includegraphics[width=0.3\linewidth]{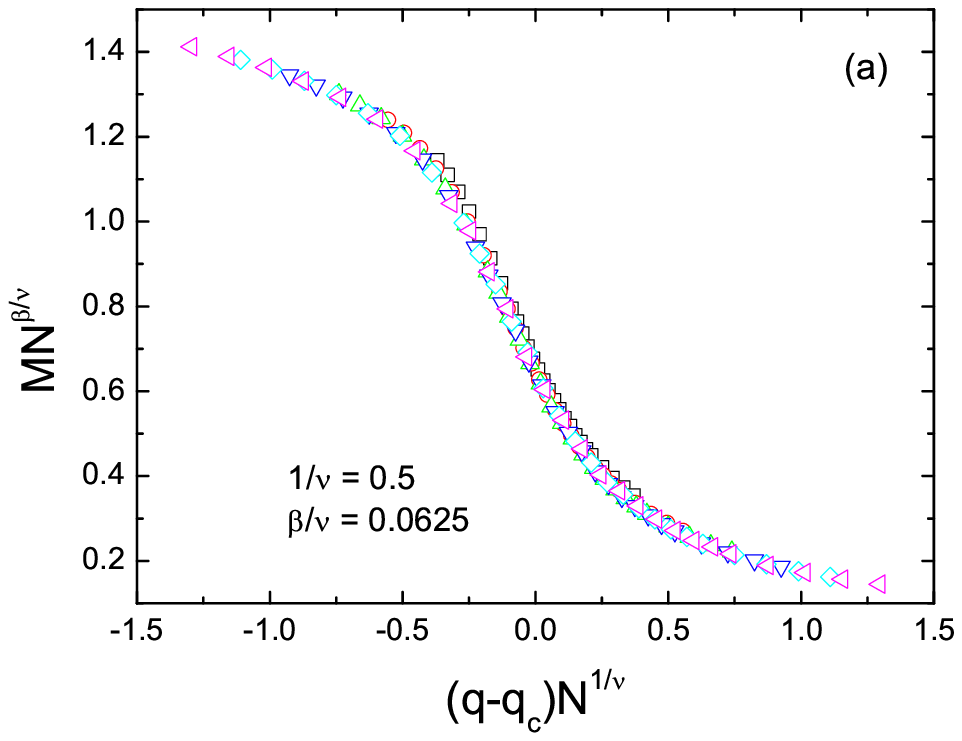}
\includegraphics[width=0.3\linewidth]{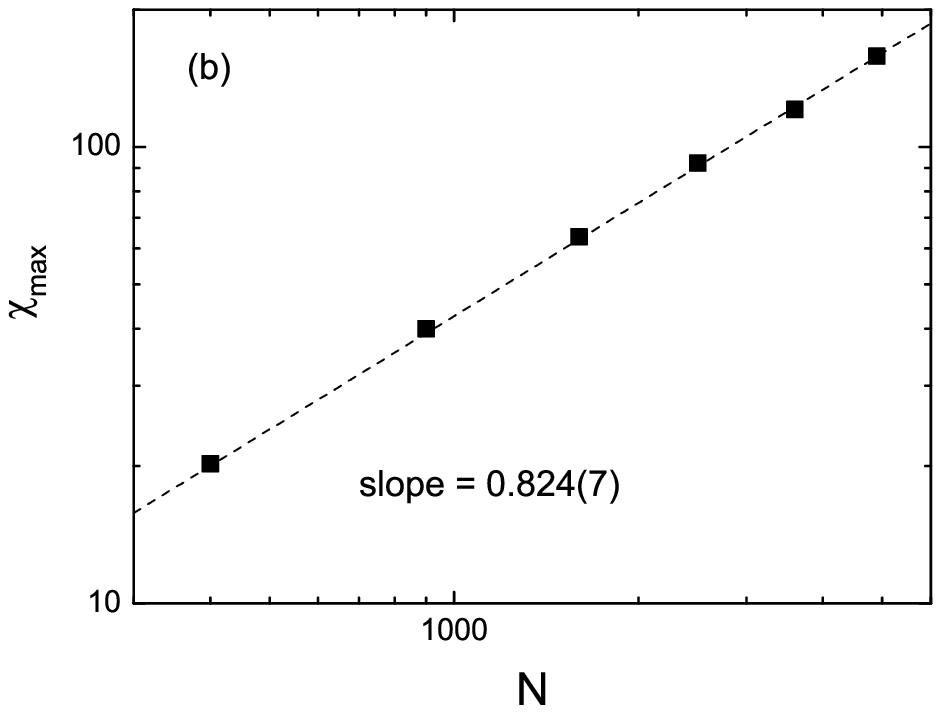}
\includegraphics[width=0.3\linewidth]{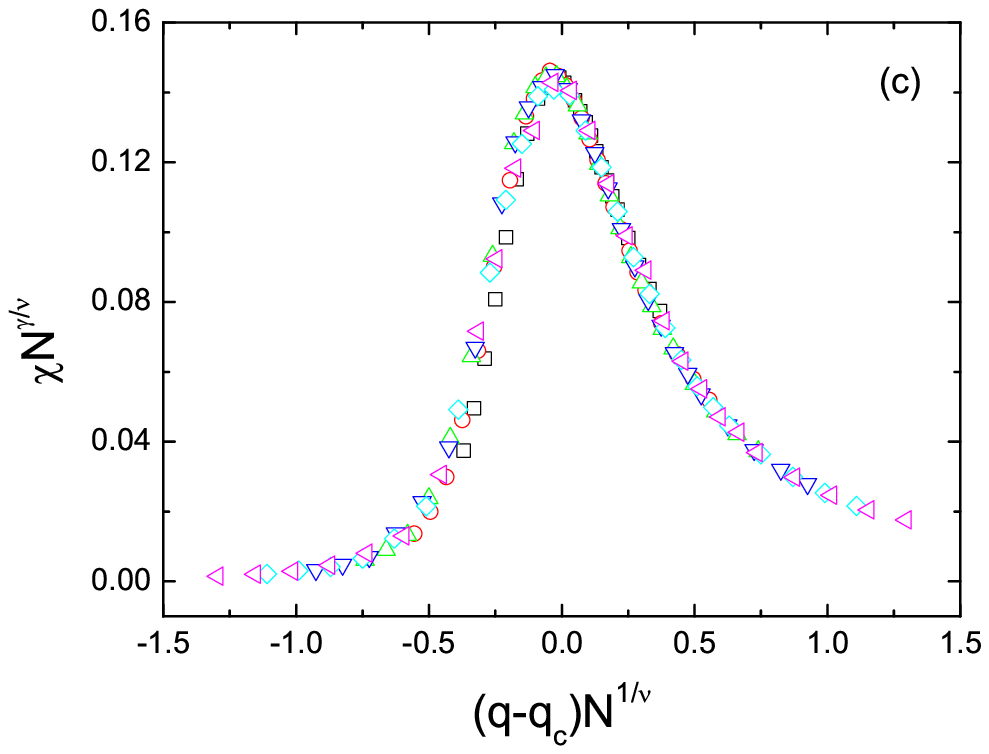}
\caption{(Color online) (a) Data collapse of the magnetization $M$
shown in Fig.~\ref{MVS}(a). The exponents used for the data
collapse are $1/\nu=0.5$ and $\beta/\nu=0.0625$. (b) Log--log plot
of the maximum of the susceptibility as a function of $N$. From it
we estimate the critical exponent $\gamma/\nu=0.824(7)$ as the
best fit of the data points. (c) Data collapse of the
susceptibility shown in Fig.~\ref{MVS}(c). The fitting exponents
are $\beta/\nu=0.0625$ and $\beta/\nu=0.825$.}\label{Scollapse}
\end{figure*}

\section{Numerical Results and Finite size scaling analysis}\label{results}
Previous investigations have shown that the majority-vote model
undergoes a phase transition from an ordered to a disordered phase
at a critical value of $q_c$. This critical value depends on the
lattice
topology~\cite{Oliveira1992jsp,Oliveira1993jpa,Marques1993jpa}. It
is worth noting that almost all previous studies assume a
 periodic boundary condition of the underlying topology. There are no
previous results for the majority-vote model on regular lattice
with a free boundary condition which is the case for the
heptagonal and dual heptagonal lattices (Fig.~\ref{HL}). Therefore
we proceed to investigate the majority-vote model on a square
lattice with free boundary condition.

\subsection{Majority-vote model on square lattice with free boundary condition}
 In Fig.~\ref{MVS} we present MC results of the majority-vote model on
square lattice with free boundary condition and plot the
magnetization, susceptibility, and reduced fourth-order cumulant
as functions of the noise parameter $q$ for several values of $N$.
As can be noticed in Fig.~\ref{MVS}(a) there is a phase transition
from an ordered state ($M>0$) to a disordered state ($M\approx0$).
In Fig.~\ref{MVS}(b), the susceptibility $\chi$ reaches a maximum
in the critical region for different system sizes which is another
typical signature for the onset of criticality. The critical point
$q_c$ can be detected from Fig.~\ref{MVS}(c), where the curves of
the reduced forth-order cumulant $U$ for different $N$ intersect
with each other. We obtain $q_c=0.074(5)$, which agrees quite well
with that for the majority-vote model on a square lattice with a
periodic boundary condition
$q_c\approx0.075$~\cite{Oliveira1992jsp,Oliveira1993jpa}.

The results of Fig.~\ref{MVS}(c) indicate that a vanishing
fraction of boundary nodes do not change the critical value of
$q$. In light of this point, we may expect the critical exponents
to be  the same. To check this assumption, we plot
$MN^{\beta/\nu}$ versus $(q-q_c)N^{1/\nu}$ in
Fig~\ref{Scollapse}(a) using the critical exponents $1/\nu=0.5$,
$\beta/\nu=0.0625$ of two dimensional Ising model. (Note that if
we measure the critical exponents in terms of linear dimension $L$
instead of $N=L^2$, then $1/\nu=1$, $\beta/\nu=0.125$.) The
excellent collapse of the curves for six different system sizes
corroborates the estimations for $q_c$ and the critical exponents
$1/\nu$ and $\beta/\nu$, and verifies the Ising universality class
of the phase transition. On the other hand, the curves for $\chi
N^{-\gamma/\nu}$ versus $(q-q_c)N^{1/\nu}$ do not overlap in the
critical region if $\gamma/\nu=0.875$ is used (results not shown),
which hints that the free boundary condition induces a strong
finite-size effect on the fluctuation of average magnetization
(the critical exponent $\gamma/\nu$). For this reason, we use
finite-size scaling by plotting $\chi_{\mathrm{max}}$ (the maximum
value of the finite-size susceptibility) as a function of $N$. We
get the slope $\gamma/\nu=0.824(7)$ of the best-fit line, as
displayed in Fig.~\ref{Scollapse}(b). The collapse of the curves
for the rescaled susceptibility verifies this estimate
[Fig.~\ref{Scollapse}(c)].

\begin{figure*}[h]
\includegraphics[width=0.3\linewidth]{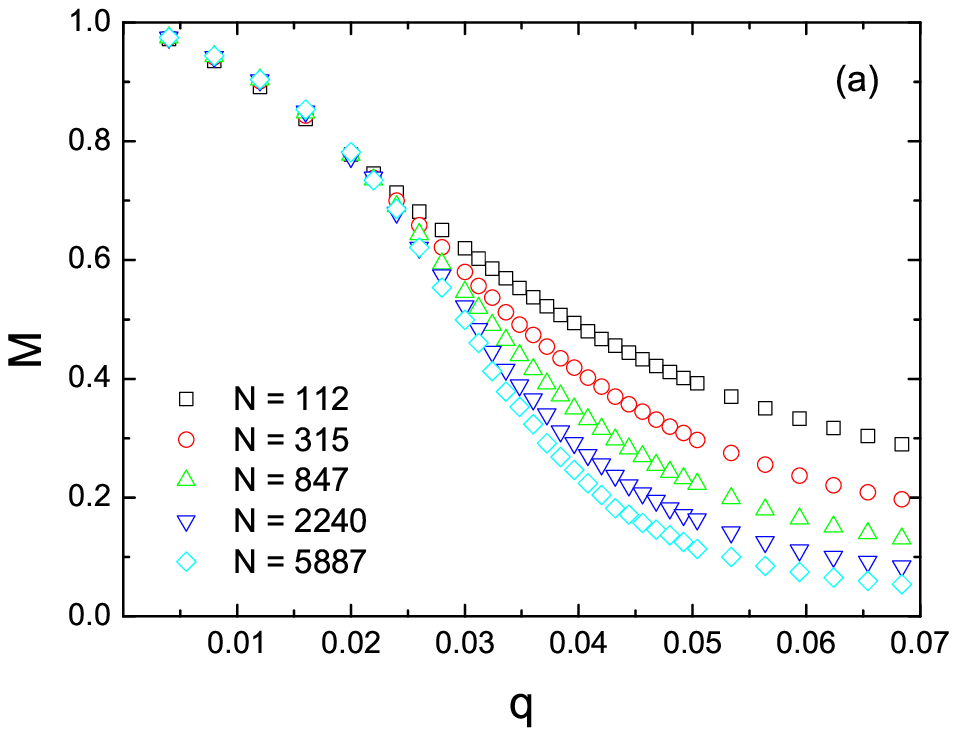}
\includegraphics[width=0.3\linewidth]{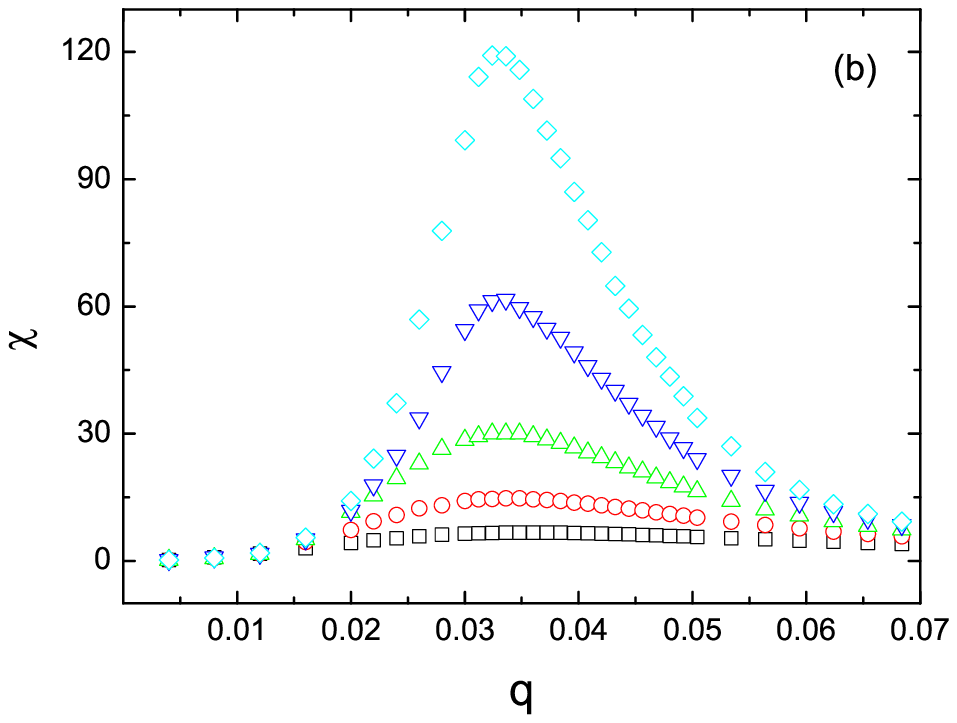}
\includegraphics[width=0.3\linewidth]{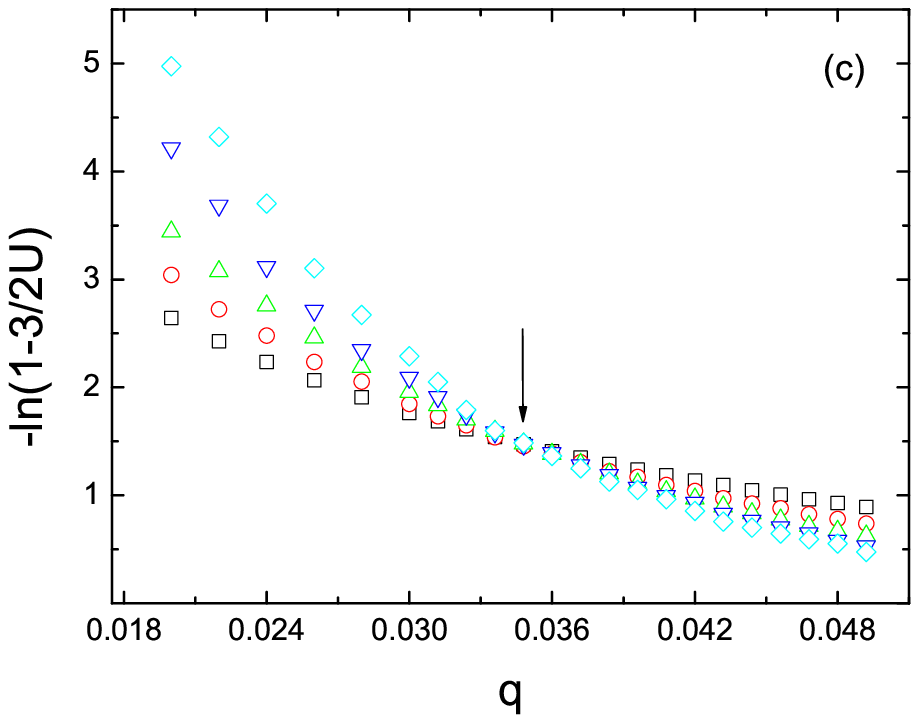}
\caption{(Color online) Majority-vote model on the heptagonal
lattices. Magnetization $M$ (a), susceptibility $\chi$ (b), and
reduced fourth-order cumulant $U$ (c), as a function of the noise
parameter $q$ for several values of the system size $N$. The
critical point $q_c=0.034(8)$ is estimated as the point at which
the different curves for different $N$ intercept each
other.}\label{MVH}
\end{figure*}

\begin{figure*}[h]
\includegraphics[width=0.24\linewidth]{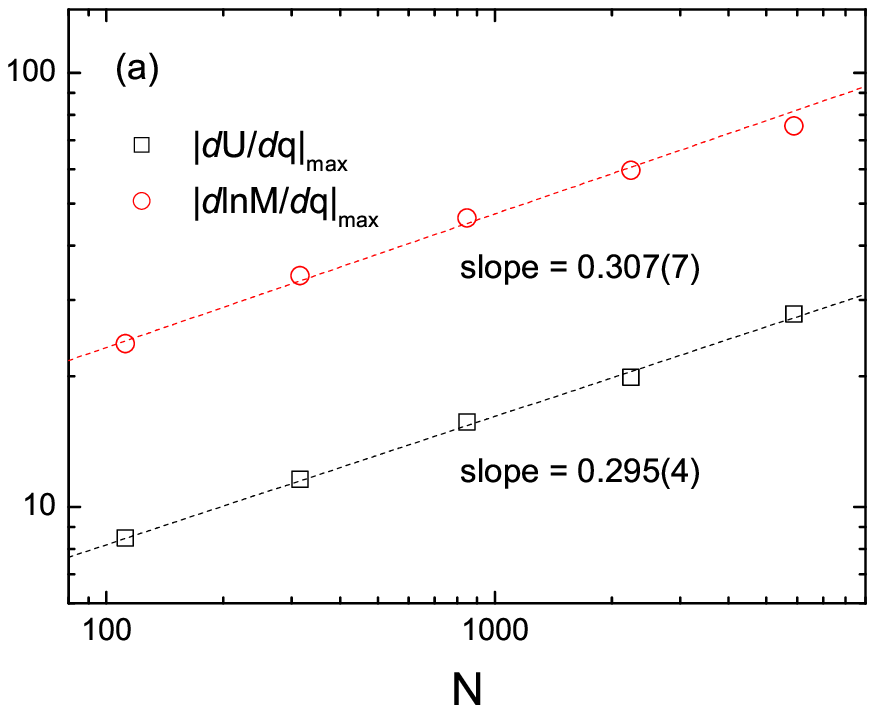}
\includegraphics[width=0.24\linewidth]{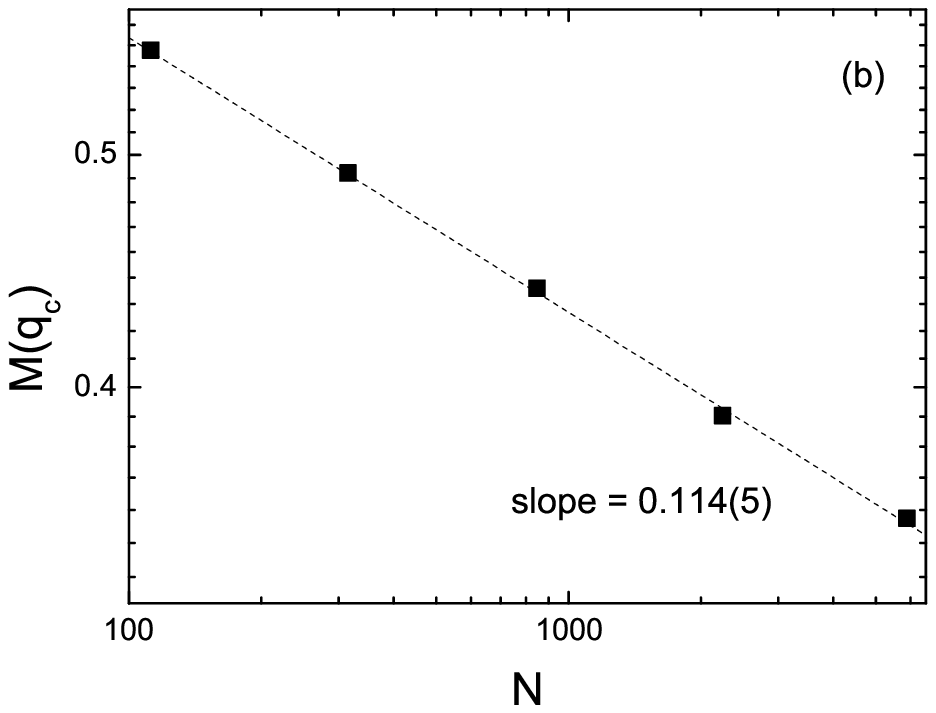}
\includegraphics[width=0.24\linewidth]{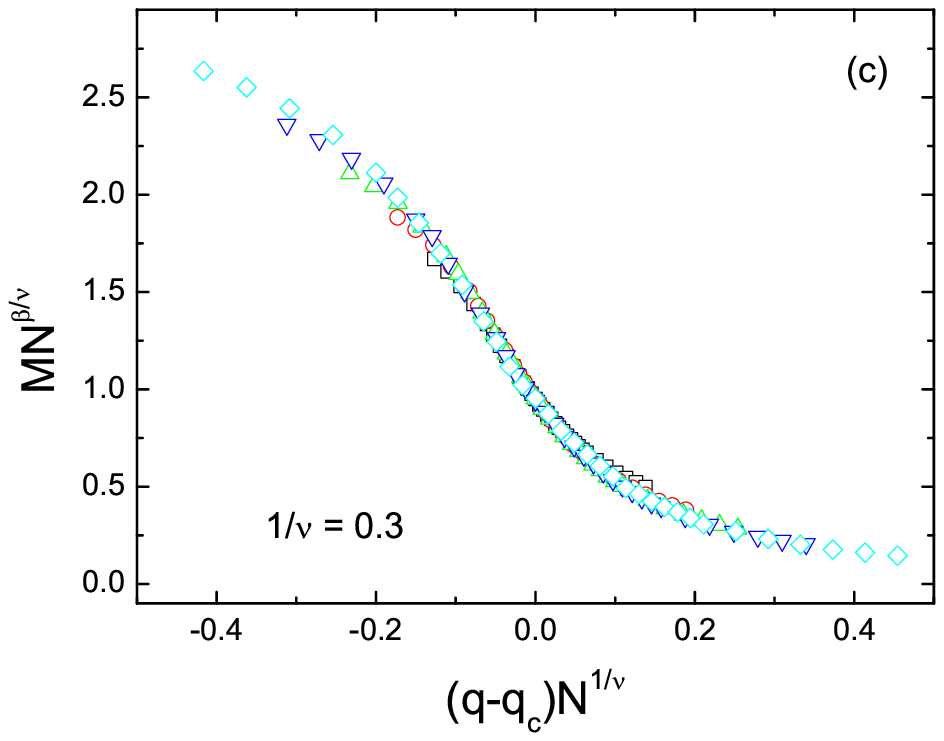}
\includegraphics[width=0.24\linewidth]{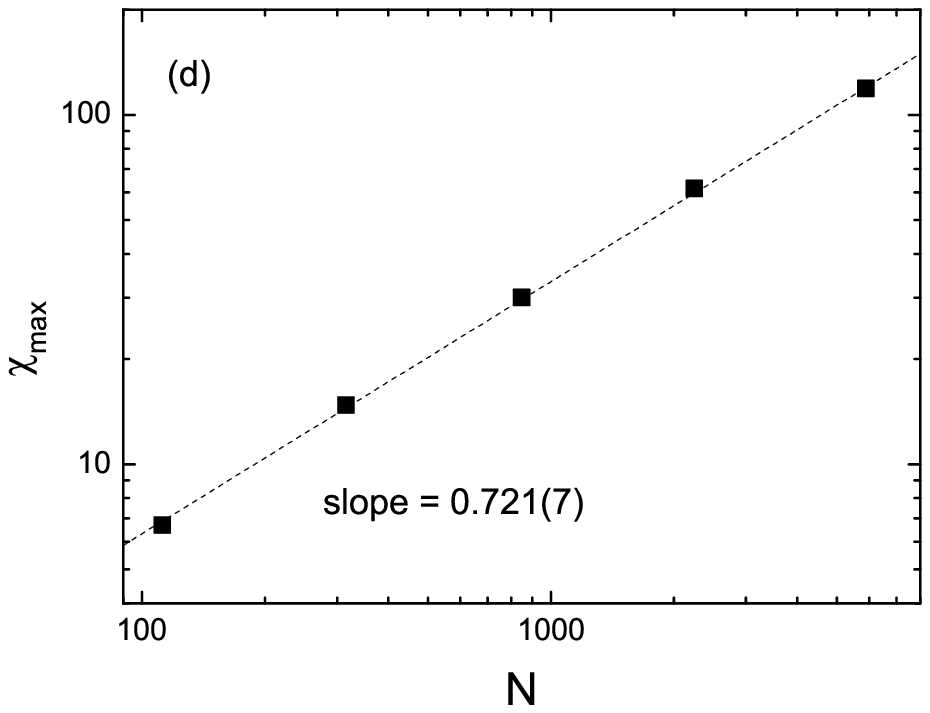}
\includegraphics[width=0.24\linewidth]{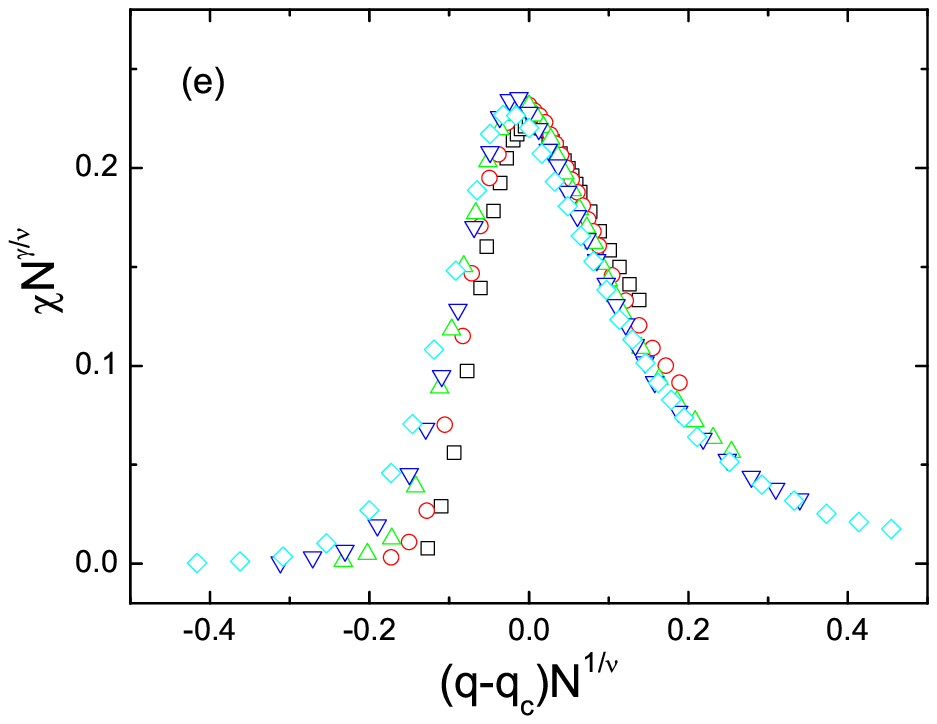}
\caption{(Color online) (a) Log--log plot of the size dependence
of the maximum values of derivatives of various thermodynamic
quantities used to determine $1/\nu$. (b)Log--log plot of the
magnetization at $q=q_c$ as a function of $N$. The slope of the
best fit  gives $\beta/\nu=0.114(5)$. (c) Data collapse of the
magnetization $M$ shown in~Fig.~\ref{MVH}(a). The exponents used
for the data collapse $1/\nu=0.3$, $\beta/\nu=0.115$. (d) Log--log
plot of the maximum of the susceptibility as a function of $N$.
From this we estimate the critical exponent $\gamma/\nu=0.721(7)$
as the best fit of data points. (e) Data collapse of the
susceptibility shown in~Fig.~\ref{MVH}(c). The fitting exponents
$\beta/\nu=0.722$, $1/\nu=0.3$.}\label{Hcollapse}
\end{figure*}

\subsection{Majority-vote model on heptagonal lattice}
We now turn our attention to the majority-vote model on the
heptagonal lattice. The simulation results for $M$, $\chi$, and
$U$ as functions of the noise parameter $q$ for different $N$ are
summarized in Fig~\ref{MVH}. As was shown in Fig.~\ref{MVS}(a),
above the critical noise level, the magnetization disappears for
larger system sizes, whereas it reaches a finite value in the
subcritical region [Fig.~\ref{MVH}(a)]. The susceptibility values
in the critical region reach their maximal values
[Fig.~\ref{MVH}(b)], and the reduced fourth-order cumulants cross
at the critical point, giving $q_c=0.034(8)$.

In order to study the universality class of the model, we proceed
to measure the critical exponents. From
Ref.~\cite{Ferrenberg1991prb}, we know that the critical exponent
$\nu$ can be obtained from the scaling behavior of certain
thermodynamic derivatives with respect to  $q$ (for example, the
derivative of the cumulant and the logarithmic derivatives of
$M$). In Fig.~\ref{Hcollapse}(a), we plot the maximum value of
these derivatives as functions of system size
 on a log--log scale.  $1/\nu$ obtained from these fits
can be seen in Fig.~\ref{Hcollapse}(a). From combining these two
estimates, we obtain $1/\nu=0.30(2)$. In Fig.~\ref{Hcollapse}(b),
we display the $N$-dependence of the magnetization at $q_c$. From
the slope of the dashed line, which corresponds to the best fit to
the data points, we estimate the corresponding value of the
critical exponent to $\beta/\nu=0.114(5)$. Using these values we
proceed to plot $MN^{\beta/\nu}$ against $(q-q_c)N^{1/\nu}$. From
the  finite-size scaling assumption that $\tilde{M}$ is a
universal function, we should, for the correct choices of $1/\nu$
and $\beta/\nu$, find a data collapse in the critical
region~\cite{Holme2006pre}. In Fig.~\ref{Hcollapse}(c) we show
that such a data collapse does indeed occur. In a similar way, we
can determine the value of $\gamma/\nu$ by fitting the data for
$\chi_{\mathrm{max}}$ as a function of $N$ in a log--log scale,
whose slope predicts $\gamma/\nu=0.721(7)$
[Fig.~\ref{Hcollapse}(d)]. By plotting $\chi N^{\gamma/\nu}$
versus $(q-q_c)N^{1/\nu}$ with $1/\nu=0.3$ and $\gamma/\nu=0.722$,
however, we only get good collapse for the curves in the
supercritical region, i.e., $q>q_c$ [Fig.~\ref{Hcollapse}(e)]. In
the subcritical region the curves deviate, suggesting an anomalous
scaling behavior.  This effect can also be seen in
Fig.~\ref{Scollapse}(d) for small system sizes. Since the boundary
vanishes with size in Fig.~\ref{Scollapse}, as does the deviation
from the scaling collapse, but the boundary does not vanish in
Fig.~\ref{Hcollapse} and neither does the deviation, we conclude
that the boundary is probably causing the anomalous scaling
behavior.

\begin{figure*}[t]
\includegraphics[width=0.3\linewidth]{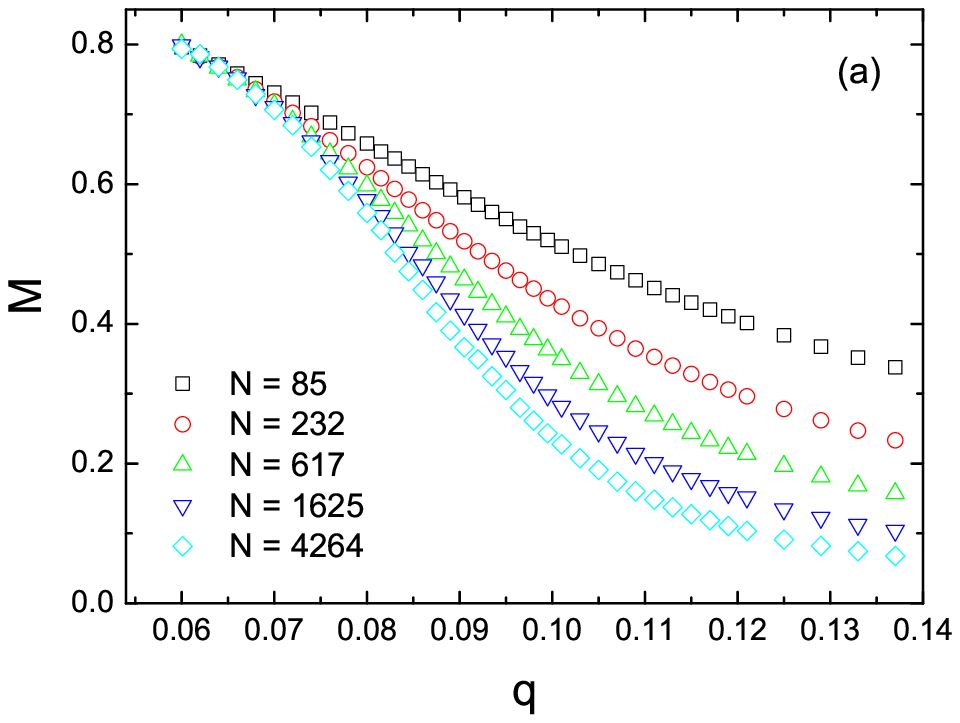}
\includegraphics[width=0.3\linewidth]{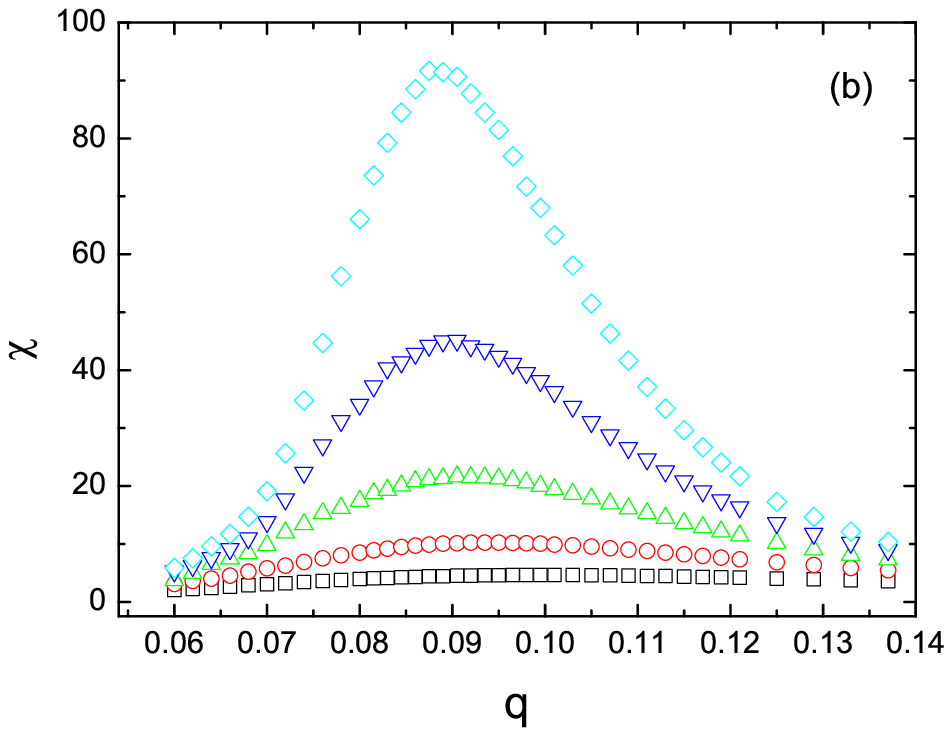}
\includegraphics[width=0.3\linewidth]{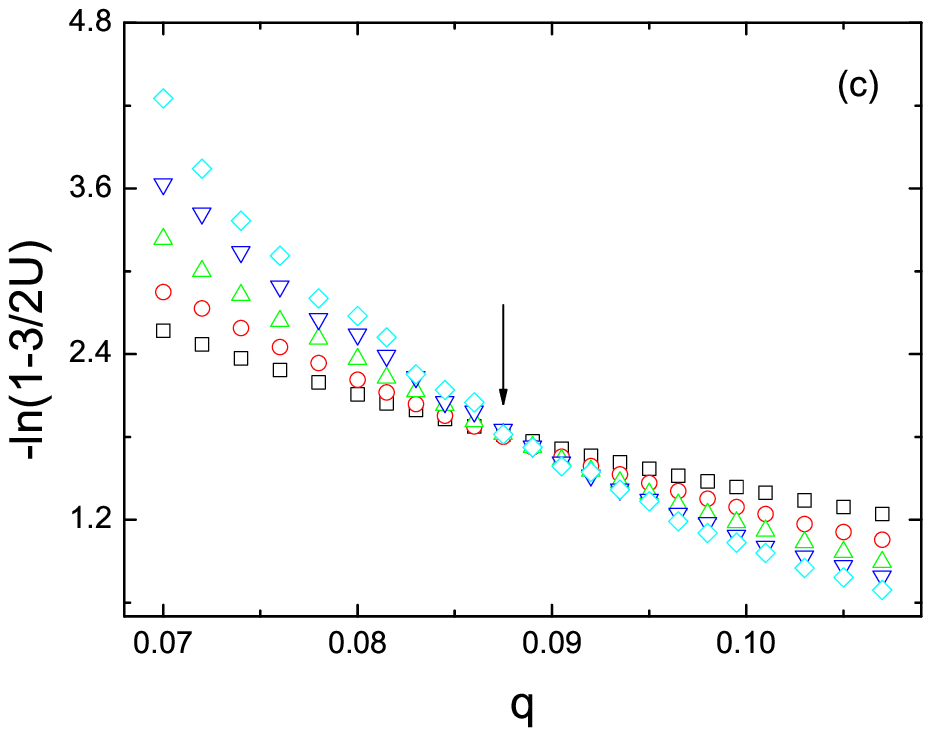}
\caption{(Color online) Majority-vote model on the dual heptagonal
lattices. Magnetization $M$ (a), susceptibility $\chi$ (b), and
reduced fourth-order cumulant $U$ (c), as a function of the noise
parameter $q$ for several values of the system size $N$. The
critical point $q_c=0.087(5)$ is estimated as the point where the
curves for different $N$ intercept.}\label{MVDH}
\end{figure*}

\subsection{Majority-vote model on dual heptagonal lattice}
\begin{figure*}[t]
\includegraphics[width=0.24\linewidth]{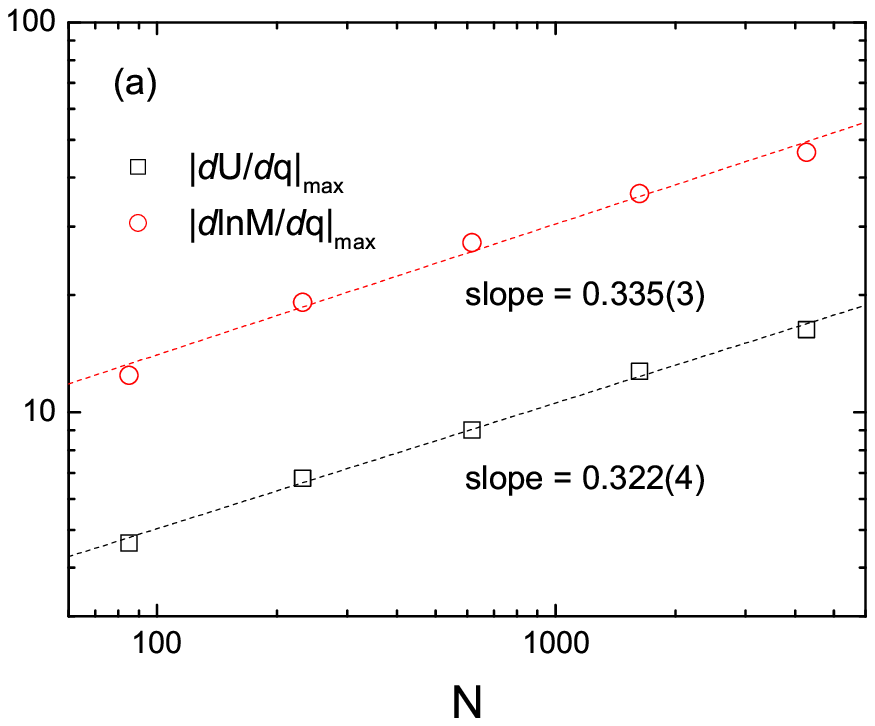}
\includegraphics[width=0.24\linewidth]{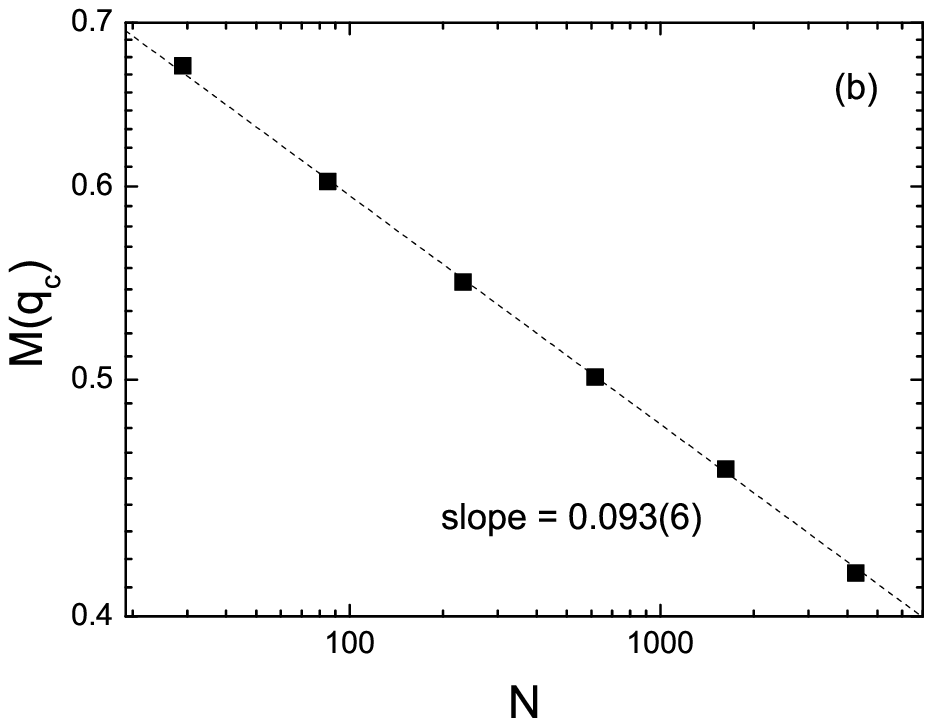}
\includegraphics[width=0.24\linewidth]{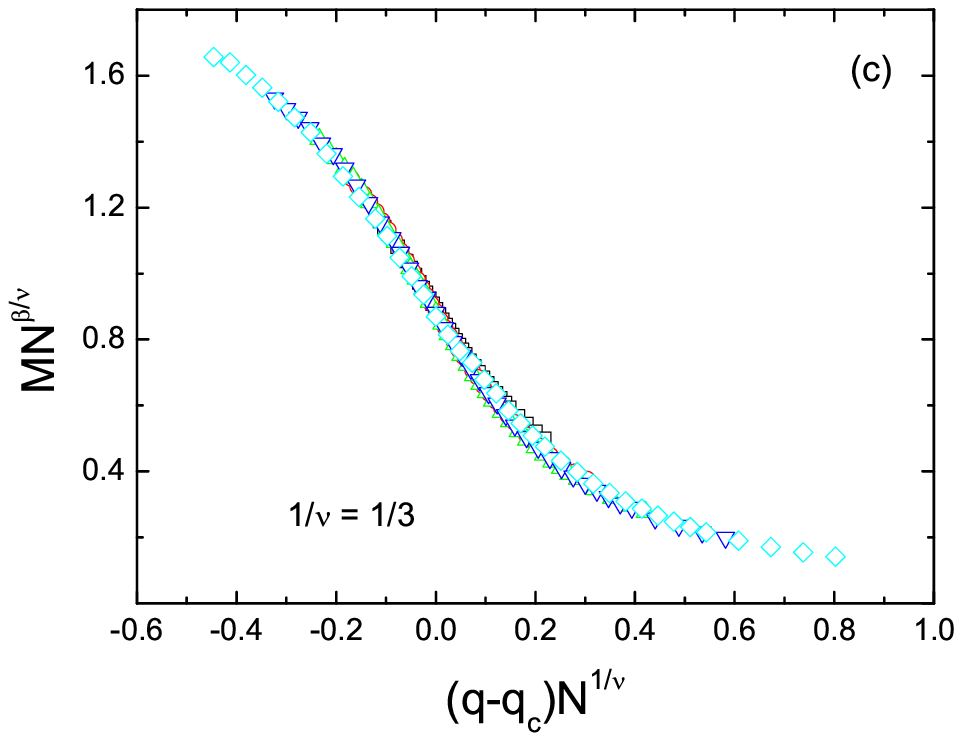}
\includegraphics[width=0.24\linewidth]{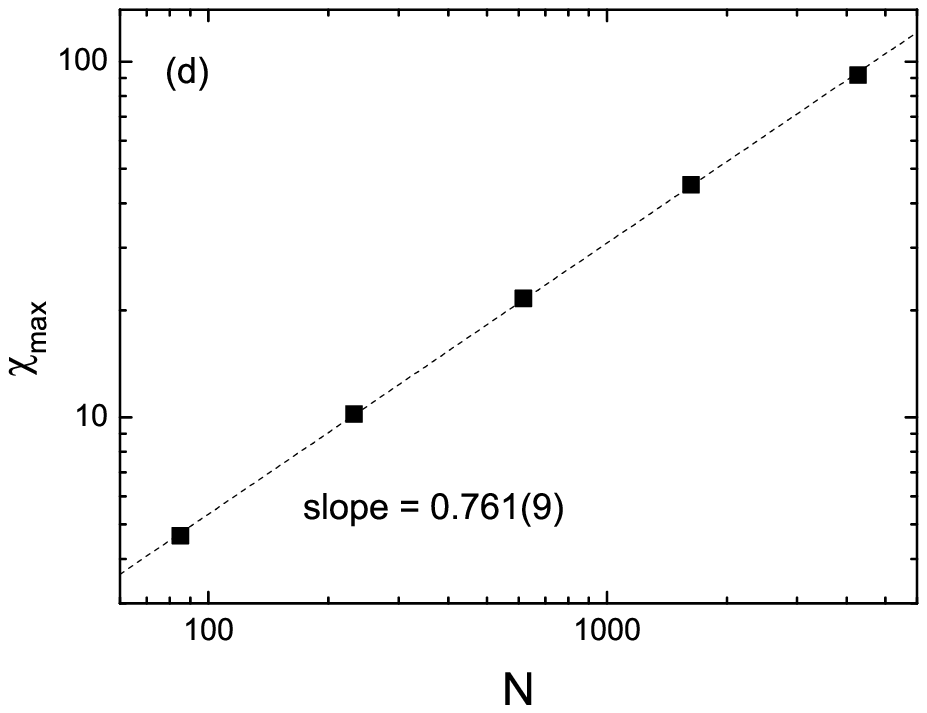}
\includegraphics[width=0.24\linewidth]{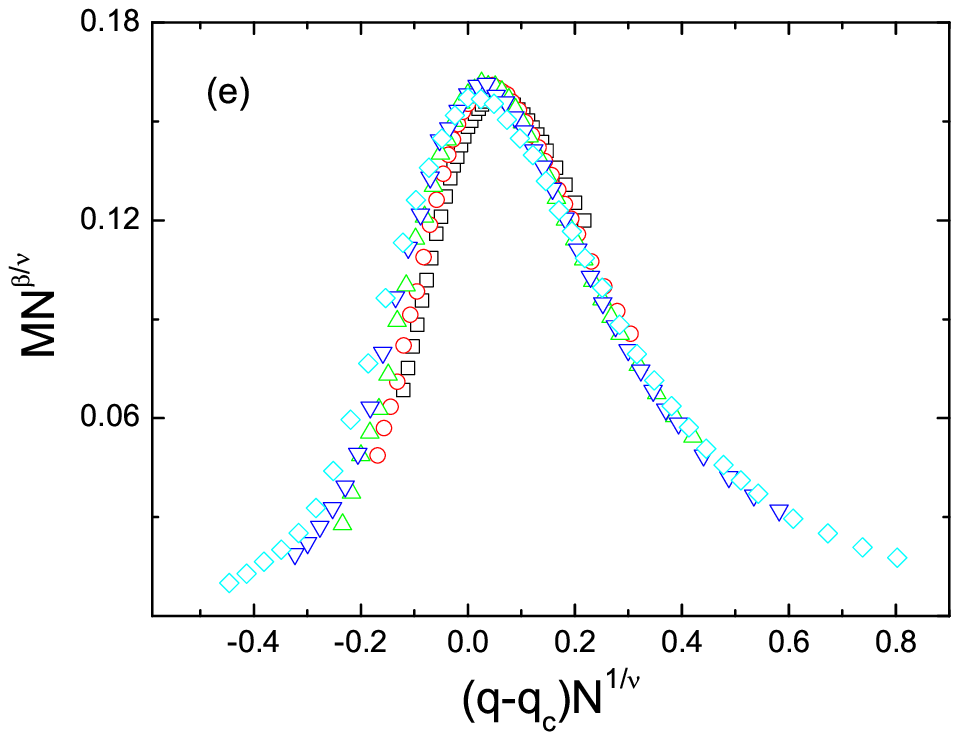}
\caption{(Color online) (a) Log--log plot of the size dependence
of the maximum values of derivatives of various thermodynamic
quantities used to determine $1/\nu$. (b) Log--log plot of the
magnetization at $q=q_c$ as a function of $N$. The slope of the
best fit of the points gives $\beta/\nu=0.093(6)$. (c) Data
collapse of the magnetization $M$ shown in~Fig.~\ref{MVDH}(a). The
exponents used for the data collapse $1/\nu=1/3$,
$\beta/\nu=0.094$. (d) Log--log plot of the maximum of the
susceptibility as a function of $N$. From this plot we estimate
the critical exponent $\gamma/\nu=0.761(9)$. (e) Data collapse of
the susceptibility shown in~Fig.~\ref{MVDH}(c). The fitting
exponents $1/\nu=1/3$, $\beta/\nu=0.762$.}\label{DHcollapse}
\end{figure*}

We  proceed to investigating the majority-vote model on the dual
heptagonal lattice. As above, we first present our numerical
results for magnetization, susceptibility and reduced fourth-order
cumulant as a function of the noise parameter (Fig.~\ref{MVDH}).
The qualitative properties of these quantities as functions of $q$
are similar to the observations in Figs.~\ref{MVS} and~\ref{MVH}.
Also here, from the intersection of the curves in
Fig.~\ref{MVDH}(c), we obtain $q_c=0.087(5)$. The critical
exponents $1/\nu$, $\beta/\nu$, and $\gamma/\nu$ are estimated in
Figs.~\ref{DHcollapse}(a), (b) and (d) to $0.32(9)$, $0.093(6)$,
and $0.761(9)$ respectively. The data collapse of the
magnetization in Fig.~\ref{DHcollapse}(c) confirms these
measurements. Also here these values do not give a good  data
collapse of $\chi$ in the region $q<q_c$.

Up to now, we obtained the critical exponents $1/\nu$, $\beta/\nu$
and $\gamma/\nu$ for the majority-vote model on the heptagonal and
dual heptagonal lattices. These exponents, and thus the
universality classes, are different from the same model on square
lattices. Interestingly, by checking the hyperscaling relation
among the exponents~\cite{Privman1990book}
\begin{equation}2\beta/\nu+\gamma/\nu=D_{\mathrm{eff}},\end{equation} we
find that $D_{\mathrm{eff}}=0.125+0.824(7)=0.949(7)$ for the
square lattice with free boundary condition,
$D_{\mathrm{eff}}=0.229+0.721(6)=0.950(6)$ for the heptagonal
lattice, and $D_{\mathrm{eff}}=0.187(2)+0.761(9)=0.949(1)$ for the
dual heptagonal lattice. Without the finite-size effects mentioned
above, these values are, we believe, consistent with $1$. (If the
scaling variable is chosen to be $L=N^{1/2}$, this means
$D_{\mathrm{eff}}=2$.) In terms of these values and taking the
finite-size scaling effect into account, we propose that our
critical exponents for the majority-vote model on negatively
curved surface also satisfy a hyperscaling relation like the
Rushbrooke and Josephson scaling laws $2\beta+\gamma=\nu
d=D_{\mathrm{eff}}$, where $d$ is a dimension of the underlying
lattice. Since the value $1/\nu$ for both the heptagonal and dual
heptagonal lattices are smaller than $0.5$, our results can be
explained by an effective dimensionality of the two hyperbolic
lattices greater than $2$. (This is reasonable---the
dimensionality of hyperbolic surfaces, embedded in an Euclidean
geometry, is also  larger than two.)

\subsection{Effect of boundary nodes}
From our numerical results, we know that the critical behaviors of
the majority-vote model on the heptagonal and dual heptagonal
lattices are different. One potential explanation comes from by
the different topological structure of the boundary nodes. (The
boundary nodes of the regular heptagonal lattice interacts less
with the inner part of the surface compared with the dual
heptagonal case.)

\begin{figure}[h]
\includegraphics[width=0.9\linewidth]{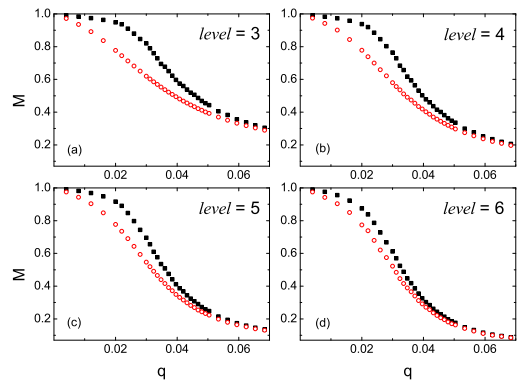}
\caption{(Color online) Magnetization $M$ as a function of the
noise parameter $q$. (a) Filled squares represent the
magnetization at the inner level $3$ of the heptagonal lattice
with  $7$ levels in total. Open circles represent the heptagonal
lattice with $3$ levels. (b), (c) and (d) show the same situation as (a) but
for different levels.}\label{innerh}
\end{figure}

\begin{figure}[h]
\includegraphics[width=0.9\linewidth]{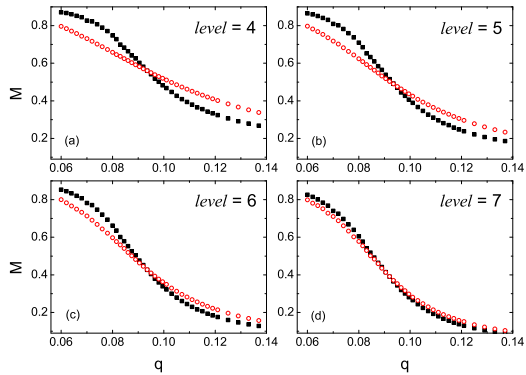}
\caption{(Color online) Magnetization $M$ as a function of the
noise parameter $q$. (a) Filled squares symbolize the
magnetization in the inner level $4$ of the dual heptagonal
lattice with $8$ levels in total. Open circles represent the
heptagonal lattice with $4$ levels. (b), (c) and (d) show  plots
corresponding to (a) for other levels.}\label{innerdh}
\end{figure}

To explore this boundary effect, one straightforward way is to
consider a (regular or dual) heptagonal lattice with a large size,
and investigate the magnetization in the center and compare that
with a heptagonal lattice of a smaller size. In particular, we
perform simulations on the heptagonal lattice at level $7$
($N=5887$), and track the magnetization in the inner part within
level $3$, $4$, $5$, and $6$ separately. Then we compare the
results with the magnetization of Fig.~\ref{MVH}(a). The same
procedure is also done for dual heptagonal lattice at level $8$
($N=4264$), but within level $4$, $5$, $6$, and $7$ respectively.
The numerical results are summarized in Figs.~\ref{innerh}
and~\ref{innerdh}. We note that the nonvanishing boundary nodes
play different roles in the ordering processes. As can be observed
in Fig.~\ref{innerh}, the magnetization in all the inner parts of
a big heptagonal lattice are greater than that on a heptagonal
lattice with same size, which indicates that the existence of a
higher level of boundary can facilitate the ordering of the inner
spins. This picture is changed, however, on the dual heptagonal
lattice. In the subcritical region $q<q_c$, the boundary nodes
promote the ordering process as before. In the supercritical
region $q>q_c$, on the contrary, they impose opposite influence on
the ordering process driving the system towards more disorder
[Fig.~\ref{innerdh}]. A structural cause for this phenomenon is,
we believe, the different local topologies of the two lattices.
Looking at Fig.~\ref{HL}, the nodes one step from the boundary
have fewer connections to the boundary in the heptagonal lattice
than its dual (the path-length going from the periphery to the
center, for systems of the same size, are longer). The boundary is
also more indirectly coupled to the interior in the heptagonal
lattice.  The boundary nodes of the dual heptagonal lattice should
thus have a stronger influence on the configuration of the
interior (we hesitate to say ``bulk properties'' since the
boundary is a finite fraction of the interior). We expect similar
stronger boundary effects for dual heptagonal lattices also exists
for other statistical spin models defined.

\subsection{Comparisons with the Ising model}
In Ref.~\cite{Oliveira1992jsp}, Oliveira found that the critical
exponents for the majority-vote model on square lattice with
periodic boundary condition are $\nu=0.99\pm0.05$,
$\gamma/\nu=1.73\pm0.05$, and $\beta/\nu=0.125\pm0.005$. These
results demonstrate the majority-vote model on a torus has the
same universal critical behavior as the equilibrium Ising model.
Recently, Yang \emph{et al.}~investigated a slightly different
majority-vote model on \emph{d}-dimensional hypercubic lattices.
Their model is identical to the original one by simple replacement
of the temperature parameter to a noise parameter like in the
present paper, $\tanh(1/kT)=(1-2q)$~\cite{Yang2008pre}. Their
simulation results suggest that the critical exponents for the
majority-vote model in three dimensions are different from those
of the Ising model, and that the results for four and five
dimensions are far from the standard mean-field values. For their
version of the majority-vote model in two dimensions, the global
and local configuration energy differences between before and
after spin-flipping are not identical~\cite{Note}, but the sign of
both energies is the same, whereas for $d\geq3$, the sign of the
two types of energy differences is not always the same. For the
Ising model, however, the differences of the global and local
configuration energy are exactly identical, regardless of
dimension. They conjectured that the discordance of the sign of
the energy difference between the global and local energy is
responsible for the different critical behaviors. Observe that our
critical exponents $1/\nu\approx0.3$ and $0.33$ in
Fig.~\ref{Hcollapse}(a) and Fig.~\ref{DHcollapse}(a), which are
smaller than $0.5$,  so from the hyperscaling relation $\nu d=1$
($d$ is the dimension) we see that  our hyperbolic lattices can be
described as having an effective dimensionality greater than $2$,
but smaller than $4$. In this sense, our present work provide
complementary support for the results obtained in
Ref.~\cite{Yang2008pre}. In another recent
work~\cite{Shima2006jpa}, Shima~\emph{et al.} studied the Ising
model on heptagonal lattices, and found the critical exponent
$\gamma/\nu=0.655$. Since our critical exponent $\gamma/\nu$, also
considering the estimated error, is quite different from this
value, we believe that the majority-vote model on the heptagonal
lattice belongs to a different universality class as the Ising
model on this topology. This result can be regarded as another
evidence for the conjecture by Yang~\textit{et
al.}~\cite{Yang2008pre} mentioned above.

\section{Conclusions}\label{conclusion}
To summarize, we have studied the critical behavior of the
majority-vote model on the heptagonal and dual heptagonal
lattices. These lattices possess a peculiar property: the ratio of
the size of the boundary to the total size  remains finite even in
the thermodynamic limit. Finite-size scaling analysis reveals that
the critical exponents for magnetization and susceptibility
deviate from those of the majority-vote model on a torus (which
belongs to the same universality class as the equilibrium Ising
model) and are also different from those of the Ising model on
heptagonal and dual heptagonal lattices. In particular, the best
fit of these exponents provided $\beta/\nu=0.114(5)$,
$\gamma/\nu=0.721(7)$ for the heptagonal lattice, and
$\beta/\nu=0.093(6)$, $\gamma/\nu=0.761(9)$ for the dual
heptagonal lattice. By comparing to the majority-vote model on
square lattices with free boundary conditions, we found that the
free boundaries result in strong finite-size scaling effect, which
in turn leads to the measured effective dimensionality smaller
than unity. Nonetheless, we believe that the critical exponents of
the majority-vote model defined on negatively curved surface also
satisfy the hyperscaling relation
$2\beta/\nu+\gamma/\nu=D_{\mathrm{eff}}$.

Furthermore, for the hyperbolic lattices, we also investigated the
effect of the boundary nodes on the ordering process. It was shown
that the boundary nodes have different functions in our two
lattices. For the heptagonal lattice, the boundary has a positive
influence on ordering, whereas for the dual heptagonal lattice the
boundary nodes can  either facilitate or inhibit ordering
depending on the magnitude of the noise. These findings are
further evidence that the underlying geometric structure determine
the critical properties of the majority-vote model. For the
future, it will be interesting  to explore the effect of boundary
nodes on the ordering processes of other statistical spin models
defined on hyperbolic lattices.

\acknowledgments{The authors wish to thank Dr.\ Seung Ki Baek for
numerous valuable comments and suggestions. Z.X.W. acknowledges
financial support from the Swedish Research Council. P.H.
acknowledges support from the Swedish Foundation for Strategic
Research, the Swedish Research Council and the WCU (World Class
University) program through the Korea Science and Engineering
Foundation funded by the Ministry of Education, Science and
Technology (R31--2008--000--10029--0).}

\bibliographystyle{h-physrev3}

\end{document}